\documentclass[journal]{IEEEtran}

\usepackage{color}
\usepackage{graphicx, subfigure}
\usepackage{amsmath}
\usepackage{cite}
\usepackage{amsfonts}
\usepackage{multirow}
\usepackage{url}
\usepackage{threeparttable}
\usepackage{rotating}
\usepackage{hyperref}
\usepackage{makecell}
\usepackage{diagbox}

\usepackage{booktabs}

\usepackage{algorithm}
\usepackage{algpseudocode}

\usepackage{tcolorbox}  

\begin{document}

\title{Prompt-Enabled Large AI Models for CSI Feedback}

\author{\normalsize {Jiajia~Guo, \IEEEmembership{\normalsize {Member,~IEEE}},
Yiming Cui, \IEEEmembership{\normalsize {Graduate Student,~IEEE}},
Chao-Kai~Wen, \IEEEmembership{\normalsize {Fellow,~IEEE}},
and Shi~Jin, \IEEEmembership{\normalsize {Fellow,~IEEE}}
}
\thanks{Jiajia~Guo, Yiming~Cui, and Shi~Jin are with the National Mobile Communications Research Laboratory, Southeast University, Nanjing, 210096, P. R. China (email: \{jiajiaguo, cuiyiming, jinshi\}@seu.edu.cn).}
\thanks{Chao-Kai Wen is with the Institute of Communications Engineering, National Sun Yat-sen University, Kaohsiung 80424, Taiwan. (e-mail: chaokai.wen@mail.nsysu.edu.tw).}
}

\maketitle

\begin{abstract}
Artificial intelligence (AI) has emerged as a promising tool for channel state information (CSI) feedback. While recent research primarily focuses on improving feedback accuracy on a specific dataset through novel architectures, the underlying mechanism of AI-based CSI feedback remains unclear.
This study explores the mechanism through analyzing performance across diverse datasets, with findings suggesting that superior feedback performance stems from AI models' strong fitting capabilities and their ability to leverage environmental knowledge. Building on these findings, we propose a prompt-enabled large AI model (LAM) for CSI feedback. The LAM employs powerful transformer blocks and is trained on extensive datasets from various scenarios. Meanwhile, the channel distribution (environmental knowledge)---represented as the mean of channel magnitude in the angular-delay domain---is incorporated as a prompt within the decoder to further enhance reconstruction quality.
Simulation results confirm that the proposed prompt-enabled LAM significantly improves feedback accuracy and generalization performance while reducing data collection requirements in new scenarios.
\end{abstract}

\begin{IEEEkeywords}
CSI feedback, large AI model, prompt engineering, channel distribution
\end{IEEEkeywords}

\IEEEpeerreviewmaketitle

\section{Introduction}

\IEEEPARstart{C}{hannel} state information (CSI) is a cornerstone of wireless communications, playing a critical role in system optimization and design \cite{10763455}. In massive multiple-input multiple-output (MIMO) systems, for example, base stations (BSs) utilize downlink CSI to design beamforming strategies and mitigate interference \cite{10375688}. However, users must transmit downlink CSI back to the BS through uplink control links, often resulting in significant overhead. This challenge becomes even more pronounced in next-generation wireless systems (e.g., extremely large-scale MIMO \cite{10379539}), which incorporate more antennas and wider bandwidths. Efficient CSI feedback methods are therefore crucial to maintaining system performance.

In current 5G systems, codebook-based methods are the predominant choice for downlink CSI feedback \cite{10375688}. However, these methods suffer from high computational complexity in codeword searches and low feedback accuracy, limiting the potential of massive MIMO systems. To address these issues, artificial intelligence (AI)-based methods have emerged as promising alternatives \cite{wen2018deep}. These methods leverage AI autoencoders, where CSI is compressed on the user side and reconstructed on the BS side \cite{guo2022overview}. Significant advancements in this area have been made to enhance feedback accuracy, including the development of advanced neural network (NN) architectures \cite{9926175} and techniques that exploit multi-domain correlations \cite{10530201}. Furthermore, joint designs combining CSI feedback with other system components, such as channel estimation and beamforming \cite{10294218}, have been shown to optimize link-level performance. 

The importance of AI-based CSI feedback is increasingly recognized in the industry. For example, the 3rd Generation Partnership Project (3GPP) has highlighted AI-enabled CSI feedback as a representative use case of AI-native air interfaces \cite{9970357}. Studies have confirmed its performance advantages over traditional methods \cite{3gpp843}. Despite their potential, AI-based CSI feedback methods face significant challenges when it comes to deployment in real-world scenarios. The requirements for large-scale data collection \cite{9714227,10622316}, sequential model training \cite{10570565}, and NN compression to address computational constraints \cite{10279462} present substantial barriers. Additionally, two key challenges remain unresolved. First, the mechanisms underlying AI-based CSI feedback are not well understood. This lack of interpretability undermines confidence in its deployment. Second, these methods face a generalization problem, where performance deteriorates significantly when inference data deviates from the training environment \cite{10288574}. This issue poses a major obstacle to their application in diverse and dynamic real-world scenarios.

Recently, large AI models (LAMs), such as ChatGPT \cite{NEURIPS2023_9f94298b,10113601,van2023chatgpt}, have achieved remarkable success in natural language processing, showcasing superior multitasking and generalization capabilities due to their extensive training and vast parameter sets \cite{kaplan2020scaling}. Inspired by these advancements, researchers in wireless communications have begun exploring the potential applications of LAMs in wireless systems. For instance, \cite{10558819} employs LAMs to build knowledge bases for semantic communications, while \cite{10579546} suggests using LAMs to develop sophisticated pre-trained wireless models. Building on these developments and addressing the challenges in AI-based CSI feedback, we pose the following question: 
\begin{center} 
\emph{Can LAMs be effectively utilized to enhance AI-based CSI feedback performance and facilitate its deployment?} 
\end{center} 
Answering this question requires addressing two primary challenges: understanding the mechanisms underlying AI-based CSI feedback and identifying the unique advantages LAMs offer in this context.

{\bf Mechanism of AI-based CSI feedback:} 
AI-based CSI feedback is often perceived as a data-driven ``black box'' due to the lack of strong theoretical foundations, making its internal workings opaque and difficult to interpret. It is well known that AI NNs learn and extract knowledge (features) by training over hundreds of epochs on substantial datasets. Most existing studies \cite[Table I]{guo2022overview} train and evaluate their proposed AI models using the publicly available CSI dataset from \cite{wen2018deep}.
While employing a standardized dataset like \cite{wen2018deep} ensures fair comparisons, it also limits insights into the dependence of AI NNs on the specific characteristics of a particular dataset. Given the data-driven nature of AI-based CSI feedback, constructing diverse CSI datasets for training and evaluating the feedback NN enables a deeper understanding of its underlying mechanisms and facilitates the identification of its key strengths and limitations.
 
{\bf Advantages of LAMs:} 
Compared to smaller AI models, LAMs provide numerous benefits. However, it is essential to identify which of these benefits are most relevant to AI-based CSI feedback. One of the most notable advantages of LAMs is their powerful fitting ability to handle multiple tasks and scenarios concurrently \cite{10113601,10367817,zheng2024large}. In practical systems, designing and training a feedback NN for each individual scenario, while achieving high performance, can be time-consuming and difficult to deploy. The exceptional multitasking capability of LAMs offers a potential solution to this issue, enabling more efficient and versatile deployment. Furthermore, prompt engineering \cite{van2023chatgpt}, a crucial and rapidly emerging technique that provides additional task-specific information to LAMs, has been extensively studied to significantly enhance LAM performance in specific tasks. Hence, if a prompt providing useful information for specific scenarios is tailored for each scenario, the feedback LAM is expected to exhibit significantly enhanced performance.

Building on these analyses, this work leverages the advantages of LAMs to enhance CSI feedback by incorporating knowledge of the mechanism underlying AI-based CSI feedback, thereby proposing a prompt-enabled LAM framework for CSI feedback. Specifically, we first investigate the mechanisms of AI-based CSI feedback by analyzing feedback accuracy across diverse CSI datasets. Key findings suggest that the superior performance of AI-based methods is attributed to their powerful fitting capabilities, which enable them to automatically handle CSI compression and reconstruction in general scenarios\footnote{In this work, general scenarios involve randomly generated channels without parameter distribution assumptions (e.g., path angle distribution), whereas specific scenarios involve channel generation based on predefined environment/distribution constraints, such as ray-tracing-based channels for a designated area.}, as well as their ability to learn and exploit environmental knowledge embedded in CSI datasets of specific scenarios. Based on these findings, we propose employing LAMs for CSI feedback, leveraging their strong fitting capabilities to achieve high-quality CSI feedback in general scenarios. Furthermore, to enhance performance, the channel distribution for each scenario---represented by the mean of the channel magnitude in the angular-delay domain---is introduced as a prompt and fed into the LAM, directly embedding environmental knowledge into the feedback process.

The main contributions of our work are as follows:
\begin{itemize}  
\item {\bf Probing AI-based CSI feedback mechanisms:} 
We conduct extensive simulations using diverse CSI datasets to explore the underlying mechanisms of AI-based CSI feedback. Our observations suggest that the powerful fitting abilities of AI-based CSI compression and reconstruction, combined with the effective learning and utilization of environmental knowledge, are key enablers behind the superior performance of AI-based CSI feedback.

\item {\bf Introducing LAMs for CSI feedback:}
We propose a novel LAM-enabled AI-based CSI feedback framework based on transformers, designed to leverage LAMs' multitasking and generalization capabilities for improving feedback accuracy in general scenarios. 

\item {\bf Incorporating environmental knowledge:}
To optimize performance in scenario-specific settings, we integrate environmental knowledge into the CSI feedback process. By representing the channel distribution as the mean of the channel magnitude in the angular-delay domain and using it as a prompt for the feedback LAM, our approach ensures that the LAM can effectively adapt to various environmental conditions. 

\item {\bf Demonstrating effectiveness:}
Simulation results demonstrate that the proposed prompt-based LAM framework consistently outperforms smaller AI models in both seen and unseen environments, achieving higher feedback accuracy while exhibiting robust generalization to unfamiliar scenarios. Additionally, our approach reduces data collection overhead and eliminates the need for online training in new scenarios, underscoring its practical advantages.
\end{itemize} 

The rest of this paper is structured as follows: Section \ref{s2Model} outlines the system model and AI-based CSI feedback framework. Section \ref{s3Observations} investigates AI-based CSI feedback mechanisms.
Section \ref{s4} introduces the proposed prompt-enabled LAM framework and elaborates on its detailed architecture.
Section \ref{s5} discusses simulation results. Finally, Section \ref{s6} provides the conclusion of this paper.

\section{System Model}
\label{s2Model}

\subsection{Signal Transmission and Channel Models}

This study considers a typical massive MIMO system where the BS is equipped with $N_{\rm t}$ uniform linear array (ULA) transmitting antennas, and the user is equipped with a single receiving antenna. The system operates in orthogonal frequency division multiplexing (OFDM) mode with $N_{\rm f}$ subcarriers and a bandwidth of $B$. The received signal on the $i$-th subcarrier, denoted as $y_i$, is expressed as
\begin{equation}  
    y_i = {\bf h}_{i} {\bf v}_{i}x_i + z_{i},   \quad  i = 1,2,\cdots, N_{\rm f},
\end{equation}
where ${\bf v}_{i} \in \mathbb{C}^{N_{\rm t} \times 1}$ represents the precoding vector designed based on the downlink CSI vector ${\bf h}_{i} \in \mathbb{C}^{1\times N_{\rm t} }$ on the $i$-th subcarrier, and $x_i\in \mathbb{C}$ and $z_{i}\in \mathbb{C}$ denote the transmitted data symbol and additive white Gaussian noise, respectively.

Adopting a path-based spatial channel model with $L$ clusters, where each cluster comprises of $L_{\rm p}$ sub-paths, the channel vector ${\bf h}_{i}$ can be expressed as in \cite{alkhateeb2019deepmimo}:
\begin{equation}
\label{scm}
    {\bf h}_{i} =\sum\limits_{l=1}^{L} \sum\limits_{p=1}^{L_{\rm p}} {{\alpha}_{l,p}}{{e}^{j({v_{l,p}}+\frac{2\pi i}{N_\text{f}}{{\tau }_{l,p}}B)}}\mathbf{a}{( \theta_{l,p})},
\end{equation}
where $\alpha_{l,p}$, $v_{l,p}$, and $\tau_{l,p}$ denote the gain, phase, and propagation delay of the $p$-th sub-path within the $l$-th cluster, respectively. 
$\theta_{l,p}$ represents the azimuth angle of departure from the BS, and $\mathbf{a}(\cdot)\in \mathbb{C}^{1\times N_{\rm t} }$ denotes the array response vector (steering vector) of the BS antenna array.
For a BS equipped with ULA transmit antennas, the steering vector $\mathbf{a}(\theta_{l,p})$ is formulated as 
\begin{equation}
\mathbf{a}(\theta_{l,p}) = \Big[1, e^{-j 2 \pi \frac{d}{\lambda} \sin(\theta_{l,p})}, \ldots, e^{-j 2 \pi \frac{(N_{\rm t} - 1)d}{\lambda} \sin(\theta_{l,p})}\Big],
\end{equation}
where $d$ and $\lambda$ represent the antenna element spacing and the carrier wavelength, respectively. In this study, the antenna element spacing $d$ is configured to be half the carrier wavelength $\lambda$.

The complete CSI matrix in the spatial-frequency domain can be obtained by stacking the channel vectors of all subcarriers as ${\bf H}' = [{\bf h}_{1}^T, {\bf h}_{2}^T, \dots,  {\bf h}_{N_{\rm f}}^T  ]^T \in \mathbb{C}^{N_{\rm f} \times N_{\rm t}} $.
Due to the limited number of scatterers, the CSI matrix exhibits sparsity in the angular-delay domain \cite{wen2018deep,guo2022overview}. It can be transformed from the original spatial-frequency domain to the sparse angular-delay domain using a two-dimensional discrete Fourier transform (DFT) as follows: 
\begin{equation}
\label{dftAD}
    {\bf  H } = {\bf F}_{\rm d}  {\bf  H }'  {\bf F}_{\rm a}\in \mathbb{C}^{N_{\rm f} \times N_{\rm t}},
\end{equation}
where ${\bf F}_{\rm d} \in \mathbb{C}^{N_{\rm f} \times N_{\rm f}} $ and ${\bf F}_{\rm d} \in \mathbb{C}^{N_{\rm t} \times N_{\rm t}}$ represent DFT matrices.
This work primarily focuses on feeding back the CSI matrix in the angular-delay domain, denoted as $\bf H$, to enhance interpretability and visualization. However, as highlighted in \cite{wen2018deep}, AI-based CSI feedback methods can also be applied to CSI feedback in the spatial-frequency domain. 

\subsection{AI-based CSI Feedback}

After obtaining the downlink CSI matrix $\bf H$ through downlink channel training, the user employs an NN-based encoder to compress the matrix and quantizes the compressed CSI using uniform or non-uniform quantizers. The compressed CSI, referred to as the codeword, is represented as the quantized codeword $\bf s$, which can be mathematically expressed as
\begin{equation}
    {\bf s} = \mathcal{Q}({\rm f}_{\rm en}({\bf H};{\bf \Theta}_{\rm en})),
\end{equation}
where ${\rm f}_{\rm en}(\cdot)$ represents the compression operation performed by the NN encoder, ${\bf \Theta}_{\rm en}$ denotes the NN parameters, and $\mathcal{Q}(\cdot)$ signifies the quantization operation.
If the CSI matrix $\bf H$ were directly transmitted back to the BS without compression and quantization, it would require transmitting $2N_{\rm f}N_{\rm t}$ real values, resulting in a substantial consumption of uplink control channel resources. Assuming a compression ratio of $\gamma<1$ and using a $b$-bit uniform quantizer to discretize the codeword, the total number of feedback bits can be expressed as 
\begin{equation}
    N_{\rm bit} = 2\gamma b N_{\rm f}N_{\rm t} .
\end{equation}
Compared to the direct feedback of the uncompressed CSI matrix $\bf H$, which requires $2N_{\rm f}N_{\rm t}$ real values, the NN-based compression method substantially reduces the feedback overhead. 

Upon receiving the feedback codeword $\bf s$ at the BS, sequential dequantization and reconstruction operations are applied to recover the downlink CSI. The recovered CSI matrix $\widehat{{\bf H}}$ is given by 
\begin{equation}
    \widehat{{\bf H}} = {\rm f}_{\rm de}( \mathcal{D}({\bf s}); {\bf \Theta}_{\rm de}  ),
\end{equation}
where  ${\rm f}_{\rm de}(\cdot)$ represents the reconstruction operation performed by the NN decoder, ${\bf \Theta}_{\rm de}$ denotes the NN parameters, and $\mathcal{D}(\cdot)$ signifies the dequantization operation.

The encoder (compression) and decoder (reconstruction) are jointly trained using an end-to-end approach. When the mean-squared error (MSE) is used as the loss function, the training process is formulated as the following optimization problem:
\begin{equation}
    \mathop{\min}\limits_{{\bf \Theta}_{\rm en} ,{\bf \Theta}_{\rm de} } \quad   \frac{1}{S} \sum_{i=1}^{S} \| \widehat{{\bf H}}_{i}-{{\bf H}}_{i} \|_{2}^{2}
\end{equation}
where $S$ denotes the number of training samples. Since the quantization operation is non-differentiable, its training gradient can be approximated as a constant, typically set to one.
In most existing studies \cite{guo2022overview}, the normalized mean squared error (NMSE) is employed as the evaluation metric, defined as
\begin{equation}
\mathrm{NMSE}=\frac{1}{S} \sum_{i=1}^{S} \frac{\|\widehat{{\bf H}}_i - {\bf H}_i\|_{2}^{2}}{\|{\bf H}_i\|_{2}^{2}}.
\end{equation}

\begin{figure}[t]
    \centering
    \includegraphics[width=0.95\linewidth]{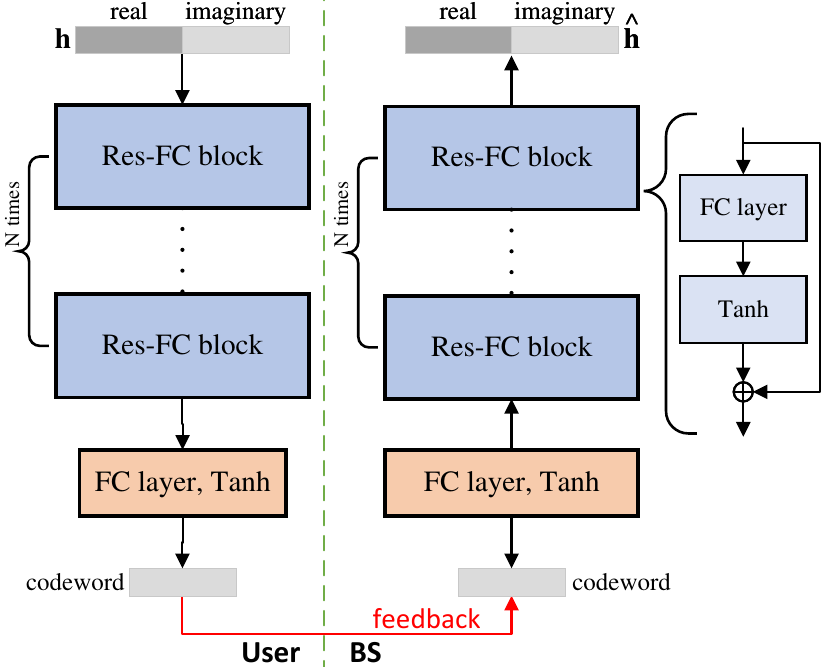}
    \caption{Detailed architecture of the FC-based CSI feedback NN, where the encoder at the user and the decoder at the BS compress and reconstruct channel vector ($\bf h$) using FC layers, respectively.}
    \label{fig:FC}
\end{figure}

\begin{figure*}[t]
 \centering
\subfigure[Cluster number $L=1$.] {
 \label{path1}
\includegraphics[width=0.46\linewidth]{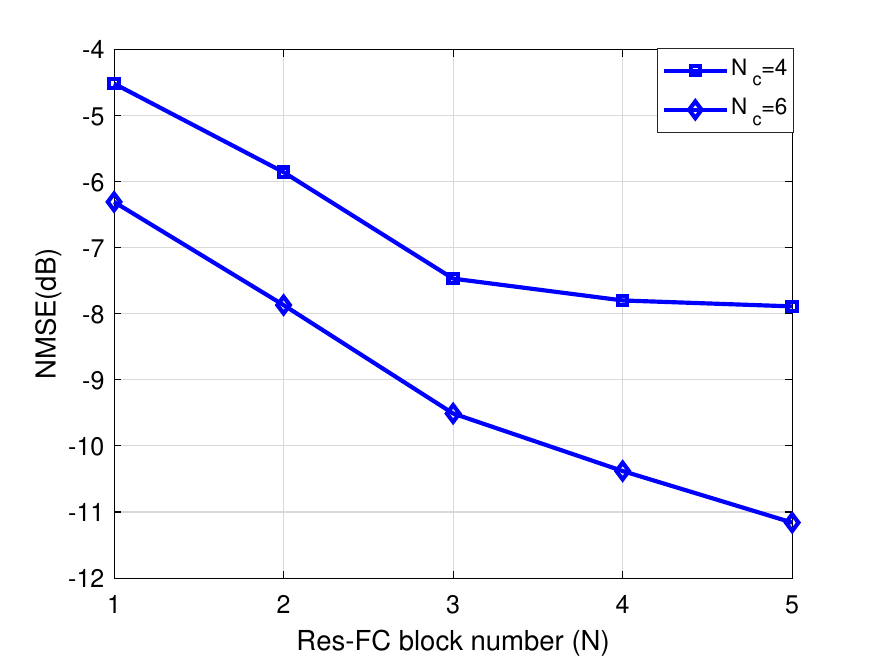}
}
\subfigure[Cluster number $L=2$.] {
\label{path2}
\includegraphics[width=0.46\linewidth]{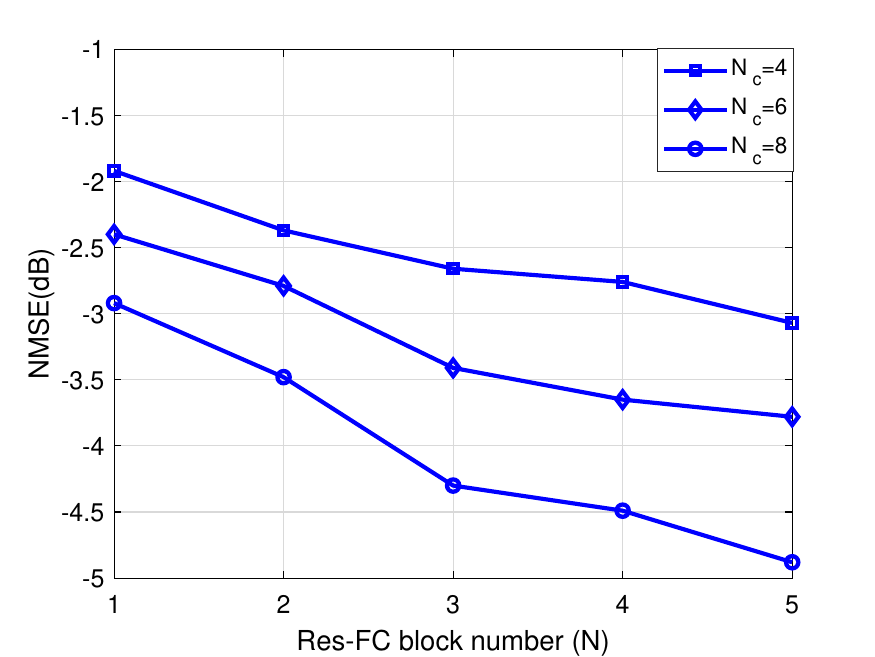}
}
\caption{NMSE performance versus the number of Res-FC blocks ($N$) for varying cluster numbers ($L$) and codeword lengths ($N_{\rm c}$). Regardless of the $N_{\rm c}$ and $L$, increasing the number of Res-FC blocks leads to improved feedback accuracy.}
\label{generalResult}
\end{figure*}
\section{Key Observations}
\label{s3Observations}

Despite numerous efforts to improve feedback accuracy, AI-based CSI feedback remains largely opaque, functioning as a ``black-box'' system with its underlying mechanisms not fully understood \cite{10476328}. This lack of transparency often compels researchers to rely on a trial-and-error approach when exploring various NN architectures to identify an optimal design for CSI feedback.

To investigate the mechanisms underlying AI-based CSI feedback, we create diverse CSI datasets and evaluate the performance of AI-based CSI feedback across these datasets. Specifically, the subsequent subsections examine two distinct scenarios. The first scenario considers general CSI conditions, where channels are stochastically generated based on predefined channel models. The second scenario focuses on specific CSI conditions, where the channel distribution is intricately tied to the surrounding physical environment. This dual-scenario analysis seeks to uncover valuable insights into the mechanisms and capabilities of AI-based CSI feedback across varying contexts.

\subsection{General Scenario}

\subsubsection{Simulation Setting}
\label{GeneralSetting}

To evaluate and analyze AI-based CSI feedback in general scenarios, channels are randomly generated in MATLAB based on the spatial channel model described in (\ref{scm}). For simplicity, we consider a narrow-band massive MIMO system, where the BS employs a ULA with $N_{\rm t}=32$ elements, and the user is equipped with a single receiving antenna. 
The number of clusters is randomly set to either 1 or 2, i.e., $L \in \{1, 2 \}$, with each cluster consisting of 5 sub-paths, i.e., $L_{\rm p} = 5$. The channel vector is converted into the sparse angular domain using a one-dimensional DFT operation and normalized within the range of $-1$ to $1$. The training, validation, and test datasets---generated based on the aforementioned channel parameters---consist of 81,000, 9,000, and 10,000 CSI samples, respectively.  

The feedback NN architecture consists of fully connected (FC) layers, excluding quantization in this section. The specific architecture is illustrated in Figure \ref{fig:FC}.
The input channel vector comprises $N_{\rm t}$ complex values, expressed as ${\bf h} \in \mathbb{C}^{1\times N_{\rm t} }$.
The real and imaginary parts are concatenated to form a ${1\times 2N_{\rm t} }$ vector,  which is then processed through $N$ residual-based FC (Res-FC) blocks. Each Res-FC block includes an FC layer with $2N_{\rm t} $ neurons, followed by a Tanh activation function, and incorporates residual learning. Subsequently, an FC layer with fewer neurons, followed by another Tanh activation function, compresses the channel vector to produce a codeword of length $N_{\rm c}$. 
At the BS, a corresponding FC layer with $2N_{\rm t} $ neurons and a Tanh activation function reconstructs the channel vector from the compressed codeword. The reconstructed vector is further refined using $N$ Res-FC blocks are subsequently employed to further enhance the reconstructed channel vector. The feedback NN is trained with a batch size of 256, 1,000 epochs, and a learning rate of 0.001.

\subsubsection{Results and Discussions}
Figure \ref{generalResult} shows the NMSE performance of the AI-based CSI feedback NN from Figure \ref{fig:FC}, plotted against the Res-FC block number ($N$) for varying cluster numbers ($L$) and codeword length ($N_{\rm c}$). In the single-cluster scenario with $L_{\rm p}=5$ sub-paths, the angular-delay domain channel vector is primarily determined by the angle and the complex gain of each sub-path. For perfect channel vector feedback, $5\times2$ real values and $5\times \log_2(N_{\rm t})=25$ bits are required for transmission of all 5 sub-paths\footnote{This assumption is an approximation and may not hold in real-world scenarios. In practice, the angles of sub-paths may not align precisely with the DFT grid points.}.

Figure \ref{path1} illustrates the feedback accuracy for the single-cluster scenario. With $N = 5$ and $N_{\rm c} = 6$, the NMSE achieves $-10$ dB, meeting the accuracy demands of real-world systems \cite{9631185}. Similar results are observed in the two-cluster scenario. Despite each cluster containing 5 sub-paths distributed within a limited angular range, the NN effectively compresses and reconstructs the channel vector with minimal overhead. This demonstrates that:  

\begin{tcolorbox}
\emph{AI-based CSI feedback can autonomously learn and leverage the inherent structure of data through end-to-end training without requiring hand-crafted design or explicit guidance.}
\end{tcolorbox}

Increasing the Res-FC block number ($N$) consistently improves feedback accuracy, regardless of the codeword length or cluster number. For instance, in the single-cluster case with $N_{\rm c}=6$ and $L=1$, the NMSE performance from $-6.31$ dB to $-10.97$ dB as $N$ increases from 1 to 5. Existing studies also reveal similar trends, where increasing NN complexity enhances its ``fitting'' capacity, leading to significant performance improvements.
For example, increasing the decoder parameters at the BS from 136K to 11.38M results in an NMSE improvement from $-3.15$ dB to $-16.86$ dB \cite[Table V]{10229094}. Smaller NNs often struggle to effectively compress and reconstruct CSI due to their limited fitting capacity to autonomously learn channel characteristics and design efficient compression algorithms. This highlights the advantage of LAMs in achieving high-precision CSI feedback, as emphasized in \cite{kaplan2020scaling}. 

The NMSE performance converges when the Res-FC block number exceeds a certain threshold, indicating an upper limit for AI-based CSI feedback performance that is challenging to precisely quantify. Notably, this study only considers scenarios with a limited number of clusters (i.e., $L=1$ and $L=2$), which are simpler compared to practical scenarios, particularly in low-frequency bands such as sub-6 GHz. More complex scenarios necessitate more powerful NNs to handle feedback effectively. Therefore: 
\begin{tcolorbox}
\emph{The utilization of LAMs is essential for maximizing the potential of AI in CSI feedback.}
\end{tcolorbox} 

\subsection{Specific Scenario}

\subsubsection{Simulation Setting}
\label{observationSimulation}
In contrast to the general scenario, where channels are randomly generated without specific assumptions about their distribution, channels in a specific scenario reflect distinct distributions influenced by the propagation environment, which are often challenging to model accurately. To evaluate and analyze AI-based CSI feedback in such specific scenarios, we generate channel datasets using the Quasi Deterministic Radio Channel Generator (QuaDRiGa) software \cite{6758357}, which produces channels based on a statistical ray-tracing model.

During channel generation, the carrier frequency is set to 2 GHz, and the 3GPP\_38.901\_UMi\_NLOS channel model \cite{3gpp2018study} is employed. For simplicity, a narrow-band system is considered. The antenna configurations for both the BS and the user are consistent with those detailed in Section \ref{GeneralSetting}. The number of clusters ($L$) is set to 4, with each cluster containing 5 sub-paths. The BS is positioned at (0 m, 0 m, 25 m), while users are randomly distributed on circles with a radius of $R$, where the distance from the circle’s center to the BS varies randomly between 20 m and 200 m.

The radii of the circles are set to 4, 5, 7, 9, and 11 m, with each circle containing 4,000 CSI samples. The training, validation, and test datasets consist of 3,150, 350, and 500 CSI samples, respectively. The NN used in this subsection follows the architecture described in Section \ref{GeneralSetting}, including the same CSI dimensions and preprocessing steps, and thus is not reiterated here.

\begin{figure}[t]
    \centering    \includegraphics[width=0.92\linewidth]{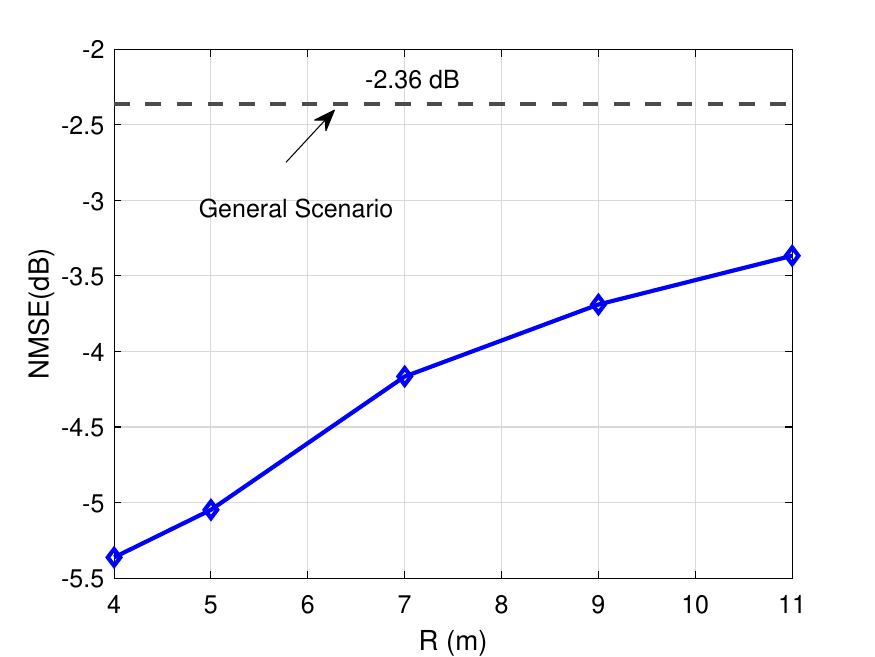}
    \caption{NMSE performance of AI-based CSI feedback versus the user range radius ($R$). The solid line represents the NMSE performance in the specific scenario generated in Section \ref{observationSimulation}, while the dashed line represents the NMSE performance in the general scenario, where channels are randomly generated using MATLAB based on spatial channel model. The cluster number ($L$),  Res-FC block number ($N$), and the codeword length ($N_{\rm c}$) are set to 4, 5, and 6, respectively.}
    \label{FCquadriga}
\end{figure}

\subsubsection{Results and Discussions}
\label{Obse2}

The cluster number ($L$),  the Res-FC block number ($N$), and the codeword length ($N_{\rm c}$) are set to 4, 5, and 6, respectively.
Figure \ref{FCquadriga} illustrates the NMSE performance of AI-based CSI feedback versus the user range radius ($R$). 
The solid line represents the NMSE performance in the specific scenario generated in Section \ref{observationSimulation}. The dashed line represents the NMSE performance in the general scenario described in Section \ref{GeneralSetting}, where channels with $L=4$  clusters are randomly generated in MATLAB based on the spatial channel model (\ref{scm}).

Firstly, regardless of the user range radius, the NMSE values for AI-based CSI feedback in specific scenarios are significantly lower than those in the general scenario. For instance, with a user range radius of 11 m, the NMSE reaches as low as $-3.37$ dB\footnote{Implementing more advanced NNs and training strategies could further enhance performance.}, representing a 20.75\% ($1$ dB) reduction in feedback errors compared to the general scenario. Similarly, the authors in \cite{10677401,10810294} highlight that NNs trained for specific scenarios can substantially enhance CSI feedback accuracy compared to generalized NNs. This observation demonstrates that:

\begin{tcolorbox}
\emph{NNs have the capability to autonomously learn environmental knowledge specific to a scenario, which is instrumental in improving AI-based CSI feedback. Leveraging such environmental knowledge is crucial for maximizing performance.}  
\end{tcolorbox}  
 
On the other hand, as shown by the solid line in Figure \ref{FCquadriga}, the performance of NNs trained for specific scenarios deteriorates as the user range radius increases. For example, reducing the radius ($R$) from 11 m to 4 m leads to a 36.61\% (2 dB) reduction in feedback errors. In compact areas, channel distributions are simpler, and environmental information plays a more dominant role in enhancing AI-based CSI feedback accuracy. A similar trend was observed in \cite{9737435}, where feedback accuracy decreased from $-12.5$ dB to $-9$ dB as the user area expands from{ 2 m $\times$ 2 m} to {6 m $\times$ 6 m}. This finding indicates that:  
\begin{tcolorbox} 
\emph{Training NNs tailored for specific areas at their smallest scale can significantly improve CSI feedback performance.}
\end{tcolorbox}

Previous studies propose a two-stage approach to leverage this characteristic: first training an NN using datasets generated randomly following the 3GPP guidelines, and subsequently fine-tuning the NN with datasets collected from specific scenarios to enhance feedback accuracy in those settings \cite{9442844,cmccHan,10381825,10508320}. While this strategy shows potential, it relies on online training, which entails significant time, computational resources, and data collection overhead. Moreover, this complex strategy poses considerable challenges to practical standardization within 3GPP \cite{9970357}. Thus, an effective strategy that eliminates the need for online training is indispensable for practical implementations.

\subsection{Summative Synopsis}

In the preceding subsections, the performance of AI-based CSI feedback has been analyzed and discussed in both general and specific scenarios, with the aim of elucidating the operational mechanisms underpinning AI-powered CSI feedback. Based on these discussions, the remarkable performance of AI in CSI feedback can be attributed to two critical factors: its powerful fitting capabilities and its ability to effectively utilize environmental knowledge.
To optimize performance and achieve the full potential of AI-driven CSI feedback, it is essential to leverage these two pivotal components to their maximum capacity. One promising approach to achieving this goal is the use of prompt-enabled LAMs, which will be introduced in detail in the next section.

\section{Prompt-enabled LAMs for CSI Feedback}
\label{s4}

\subsection{Motivation}
\label{motivation}

Building on the insights from the previous section, our goal is to design a robust NN capable of excelling in two critical aspects: fitting complex CSI compression and reconstruction tasks, and effectively extracting and utilizing environmental knowledge specific to a given scenario.

\subsubsection{LAM---Powerful Fitting Capabilities} 
The fitting capabilities of NNs, which automatically handle
CSI compression and reconstruction,  improve with increased complexity, including greater depth and width \cite{6114ICML,behler2021four}. This is because deeper and wider NNs possess more parameters, enabling them to capture intricate relationships within data. However, overparameterized NNs with a large number of parameters are prone to overfitting. According to the scaling laws proposed in \cite{kaplan2020scaling,pnas2311878121}, larger models trained on substantial datasets often exhibit superior performance and are less susceptible to overfitting. Consequently, LAMs and sufficiently large training datasets are essential for achieving powerful fitting capabilities in AI-based CSI feedback.

Despite the presence of LAMs, there is often a trade-off between generalization and performance \cite{PACE}. A highly generalized NN may struggle to perform optimally in specific scenarios compared to an NN trained specifically for that scenario. This highlights the importance of integrating environmental knowledge into LAMs to fully maximize their potential.

\subsubsection{Prompt---Environmental Knowledge Utilization} 
Effectively utilizing environmental knowledge for CSI feedback involves addressing two key questions: how to define environmental knowledge, and how to integrate it into LAMs.

\paragraph{Environmental Knowledge Definition}

In existing studies \cite{10677401,9737435,9930135,AIMLSystems,10465185,10180075}, environmental knowledge is often left undefined, implicitly extracted and utilized through training. Feedback NNs are typically trained on CSI data from specific scenarios, leading to optimal performance within those scenarios but significantly poorer performance in others.

\begin{figure}[t]
    \centering
    \includegraphics[width=1\linewidth]{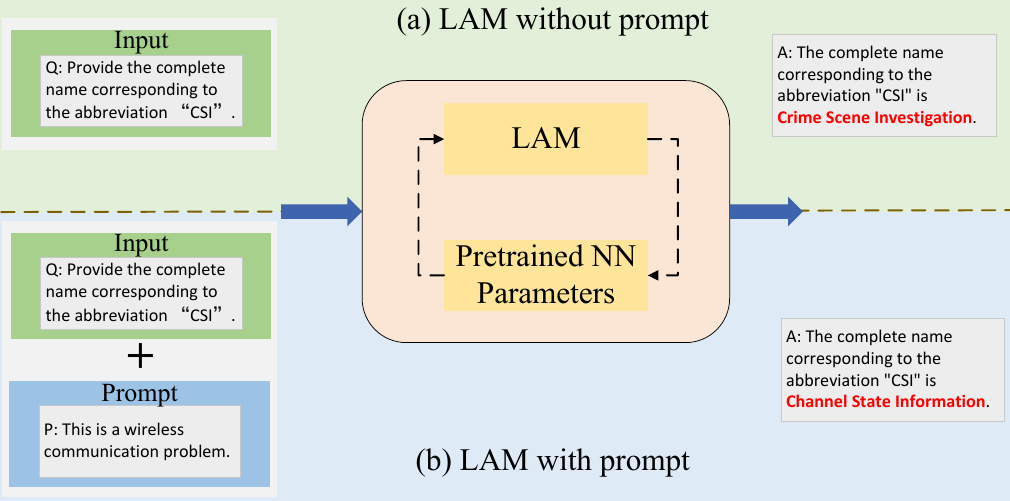}
    \caption{Illustration of LAMs without and with prompts. In this illustration, the application of prompts, specifically in specifying question types, directs LAMs toward the intended outcomes.}
    \label{LAMprompt}
\end{figure}

To address this gap, we categorize environmental knowledge into two distinct types: direct and indirect.
\begin{itemize}
    \item Direct Environmental Knowledge: The physical propagation environment directly affects signal transmission and can be considered as direct environmental knowledge. This includes information critical for CSI feedback, such as the floor plan \cite{10620428}. However, direct environmental knowledge often includes irrelevant data, is challenging to model accurately, and may raise privacy concerns.

    \item Indirect Environmental Knowledge: Indirect environmental knowledge refers to the channel distribution influenced by the physical propagation environment. This includes information like channel model types, path distributions, and other statistical characteristics. Although existing studies \cite{10677401,9737435,9930135,AIMLSystems,10465185,10180075} do not explicitly define or extract environmental knowledge, they leverage channel distribution patterns learned from training data to enhance feedback accuracy.
    
\end{itemize}  

Recent research has explored representations of indirect environmental knowledge. For example, \cite{cmccHan} extracts stochastic parameters from channel data to analyze wireless propagation characteristics, while \cite{baur2024evaluation} introduces a codebook fingerprint---a histogram of codewords depicting channel directional characteristics. Both studies highlight the importance of path-related information, such as direction, delay, and other factors, as critical components of the channel distribution. However, their primary focus is on utilizing this information to facilitate channel data collection and reduce associated overheads. They largely overlook the potential of this path-related information to significantly enhance CSI feedback accuracy and eliminate the need for online training.


\paragraph{Environmental Knowledge Incorporation}
Traditionally, feedback NNs are trained separately for each scenario, implicitly leveraging environmental knowledge to enhance accuracy. However, this approach incurs significant overhead, including computational resources, data collection, and NN parameter transmission, making deployment challenging.

Recently, prompt engineering has gained significant attention in the context of LAMs \cite{van2023chatgpt}. This technique employs task-specific prompts to optimize model performance without requiring modifications to the underlying model architecture. As shown in Figure \ref{LAMprompt}, prompts can guide LAMs to better align with specific tasks. For example, when tasked with expanding abbreviations, a domain-specific prompt (e.g., ``wireless communication'') can guide the LAM to correctly interpret ``CSI'' as ``channel state information'' rather than the unrelated ``crime scene investigation.'' This simple yet effective approach enables LAMs to tailor their output to specific contexts, significantly improving accuracy.

Inspired by the success of prompt engineering, we propose treating CSI feedback for different scenarios as distinct tasks, as demonstrated in \cite{9930135,AIMLSystems}. By integrating prompts that describe scenario-specific environmental knowledge, LAMs can effectively adapt to unique characteristics and improve CSI feedback accuracy without requiring online training or additional model reconfiguration.

\begin{figure}
 \centering
\subfigure[Offline training] {
 \label{frameworkOffline}
\includegraphics[width=1\linewidth]{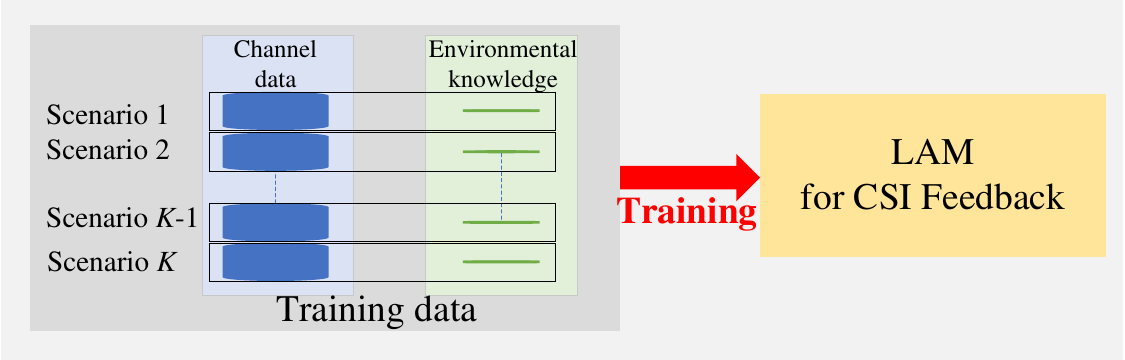}}
\subfigure[Online inference]{
\label{frameworkOnline}
\includegraphics[width=1\linewidth]{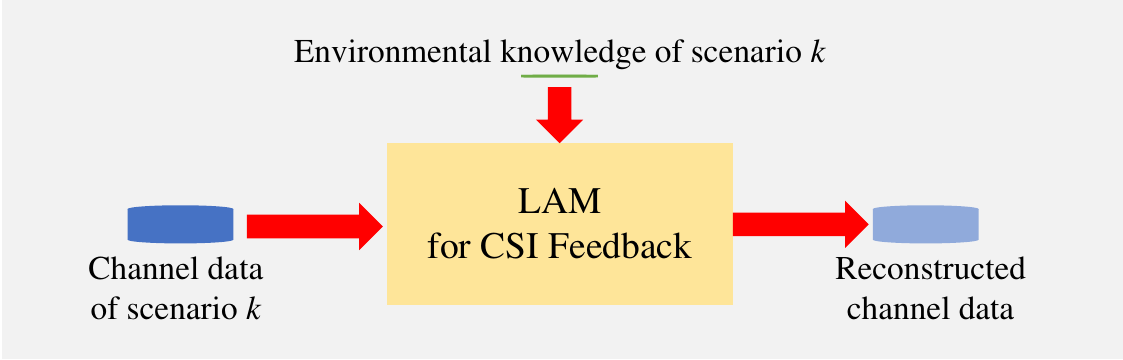}}
\caption{Illustration of the general framework of prompt-enabled LAMs for CSI feedback, including offline training and online inference.}
\label{frameworkLAMprompt}
\end{figure}

\subsection{Main Framework}

Figure \ref{frameworkLAMprompt} illustrates the general framework of prompt-enabled LAMs for CSI feedback, encompassing both offline training and online inference stages. Unlike conventional AI-based CSI feedback systems that rely solely on channel data, this approach integrates both channel data and environmental knowledge throughout the training and inference processes. 
Specifically, during NN training, a dataset comprising extensive channel data from $K\gg 1$ scenarios and $K$ categories of environmental knowledge is used to train the LAM for CSI feedback. In the inference stage, channel data is input into the LAM alongside a prompt (the corresponding environmental knowledge for the scenario). This approach eliminates the need for online training when encountering new scenarios while effectively leveraging scenario-specific environmental knowledge.

\begin{figure}[t]
    \centering
    \includegraphics[width=0.99\linewidth]{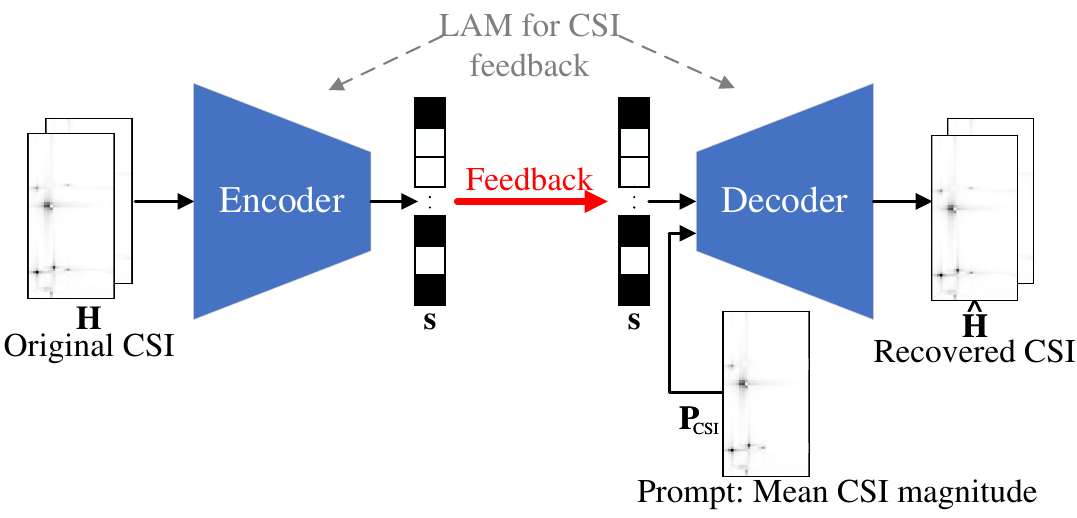}
    \caption{Framework of prompt-enabled LAMs for CSI feedback. The mean CSI magnitude, which describes the channel distribution, serves as the environmental knowledge (prompt) to facilitate CSI reconstruction at the BS using the feedback codeword ($\bf s$).}
    \label{frameworkFeedback}
\end{figure}

Figure \ref{frameworkFeedback} presents the detailed framework of the proposed approach. An AI-based autoencoder is employed to achieve CSI feedback. To enhance NN fitting capabilities, both the encoder for CSI compression and the decoder for CSI reconstruction are composed of powerful NN blocks, which are detailed in the following subsection. This subsection focuses on the design and integration of prompts (environmental knowledge) and the associated data collection workflow.

\subsubsection{Prompt (Environmental Knowledge) Design}
As discussed in Section \ref{motivation}, the channel distribution can serve as a prompt that characterizes the feedback scenario. Inspired by \cite{cmccHan}, the distribution of path parameters could theoretically represent channel distribution. However, extracting path parameters from individual channel samples using methods like the orthogonal matching pursuit (OMP) algorithm is computationally intensive and unnecessary for this purpose.

Instead of aiming to accurately identify each path and its parameters, the goal is to estimate the probability distribution of path angles and delays. In the angular-delay domain, as shown in (\ref{dftAD}), the CSI matrix comprises complex values. However, only the magnitude of the CSI, not the phase, contains information about path angles and delays. Thus, this study adopts the mean magnitude of all downlink CSI samples as the environmental knowledge (prompt). This metric, denoted as ${\bf P}_{\rm CSI}$, can represent the channel distribution by approximating the probabilities of path angles and delays. 
Beyond this prompt, various other prompts can also be effectively utilized. Given the aim of this study is not to maximize feedback LAMs' performance, a simple prompt, ${\bf P}_{\rm CSI}$, is used. Employing superior prompts, a topic reserved for future research, has the potential to notably boost performance.


\subsubsection{Prompt (Environmental Knowledge) Integration}
As shown in Figure \ref{LAMprompt}(b), systems like ChatGPT \cite{10113601,NEURIPS2023_9f94298b} integrate prompts with input data to guide LAMs in producing task-specific outputs \cite{van2023chatgpt}. However, CSI feedback involves a two-sided autoencoder structure, where the encoder at the user compresses CSI, and the decoder at the BS reconstructs it.

While incorporating prompts at both the encoder (user) and decoder (BS) could maximize prompt benefits, doing so poses significant challenges. First, when a user enters a new scenario, the BS must transmit pre-generated prompts (environmental knowledge) to the user, consuming valuable downlink resources, especially for users at the scenario's edge. Second, adding computational complexity at the user is impractical, as user devices typically have limited processing power compared to the BS.

To address these issues, this study uses prompts exclusively at the decoder (BS), as illustrated in Figure \ref{frameworkFeedback}. Once the BS receives the feedback codeword ($\bf s$), it combines it with the pre-generated prompt (${\bf P}_{\rm CSI}$) before feeding it to the decoder. This setup enables the decoder to leverage both the feedback codeword and environmental knowledge (prompt) for CSI reconstruction, improving accuracy without requiring user intervention. The prompt selection, which could be based on the user’s position, is left for future investigation.
 
\begin{figure}[t]
    \centering
    \includegraphics[width=0.99\linewidth]{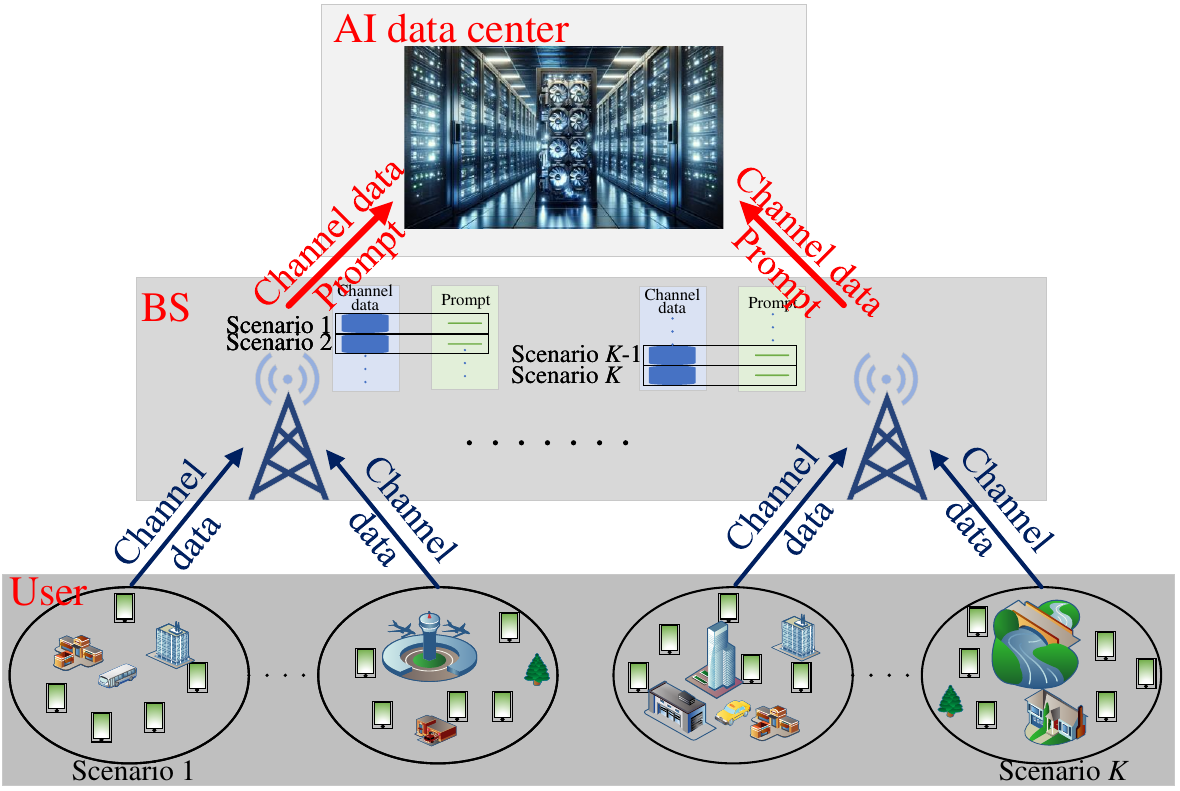}
    \caption{Illustration of data collection for LAM training. During idle periods, users send high-quality channel data to the BS, which forwards the data and corresponding prompts (${ \bf P }_{ \rm CSI }$) to an AI data center for training the CSI feedback LAM.}
    \label{TrainingDataCollect}
\end{figure}

\subsubsection{Data Collection for Training and Inference}
The proposed prompt-enabled LAM for CSI feedback requires two types of data: channel data and prompts.

\paragraph{Training} 
Unlike previous studies that collected CSI data without categorization, the training datasets in this work are organized into $K$ distinct scenarios, as previously defined. Dividing data into distinct scenarios presents significant challenges. For instance, in \cite{9737435}, different offices are treated as separate scenarios, whereas in \cite{10677401,10810294}, users are partitioned based on their positions using clustering techniques derived from position or CSI data. 
In this work, scenario categorization is not considered; instead, CSI data is directly generated for $K$  scenarios. Further details on the data generation process are provided in Section \ref{CSIgenera}.

During the training phase, as illustrated in Figure \ref{TrainingDataCollect}, users of all $K$ scenarios transmit high-quality channel data to their corresponding BS during idle periods. 
The BS calculates the mean CSI magnitude (${\bf P}_{\rm CSI}$) for each scenario and forwards both the channel data and the corresponding prompt (mean CSI magnitude) to a centralized data center. The data center aggregates channel data and prompts from numerous BSs across all $K$ scenarios to train the CSI feedback LAM, as shown in Figure \ref{frameworkFeedback}.

Once training is complete, the trained encoder and decoder of the CSI feedback LAM are deployed to the user and the BS, respectively. Importantly, the feedback LAM is designed not only to perform well in scenarios present in the training dataset but also to generalize effectively to unseen scenarios.
    
\paragraph{Inference}
In contrast to the training phase, the inference phase requires only the prompt, specifically the mean CSI magnitude (${\bf P}_{\rm CSI}$) in this study. For scenarios included in the training dataset, the prompt can be directly retrieved without incurring additional overhead. 
 
For new or unseen scenarios, only the prompt needs to be collected, eliminating the need to gather comprehensive channel data. This represents a significant reduction in data collection overhead compared to current online training approaches \cite{guo2022overview}. Assuming there are $U$ users in a given scenario contributing to prompt collection, with each user collecting $N_{m}$ CSI samples, the mean CSI magnitude ${\bf P}_{\rm CSI}$ can be calculated as
    \begin{equation}
\label{CSImagnitude}
{\bf P}_{\rm CSI} = \frac{1}{\sum_{m=1}^{U}{N_m}} \sum_{m=1}^{U} \sum_{n=1}^{N_{m}} |{\bf H}_{n}^{m}|,
\end{equation}
where ${\bf H}_{n}^{m}$ represents the $n$-th channel collected by the $m$-th user. 

To construct the prompt (${\bf P}_{\rm CSI}$), each user sends the sum of the magnitudes of their collected CSI matrices ($\sum_{n=1}^{N_{m}} |{\bf H}_{n}^{m}|$) and the corresponding sample count ($N_m$) to the BS. The transmission overhead for this operation is minimal, as it is equivalent to sending a single CSI magnitude sample along with the sample count, far less than transmitting $N_m$ complex CSI samples.
Using this information, the BS computes the mean CSI magnitude (${\bf P}_{\rm CSI}$) by applying (\ref{CSImagnitude}), leveraging the received sum of channel magnitudes and sample counts from all users in the scenario.

 \begin{figure*}[t]
    \centering
    \includegraphics[width=0.75\linewidth]{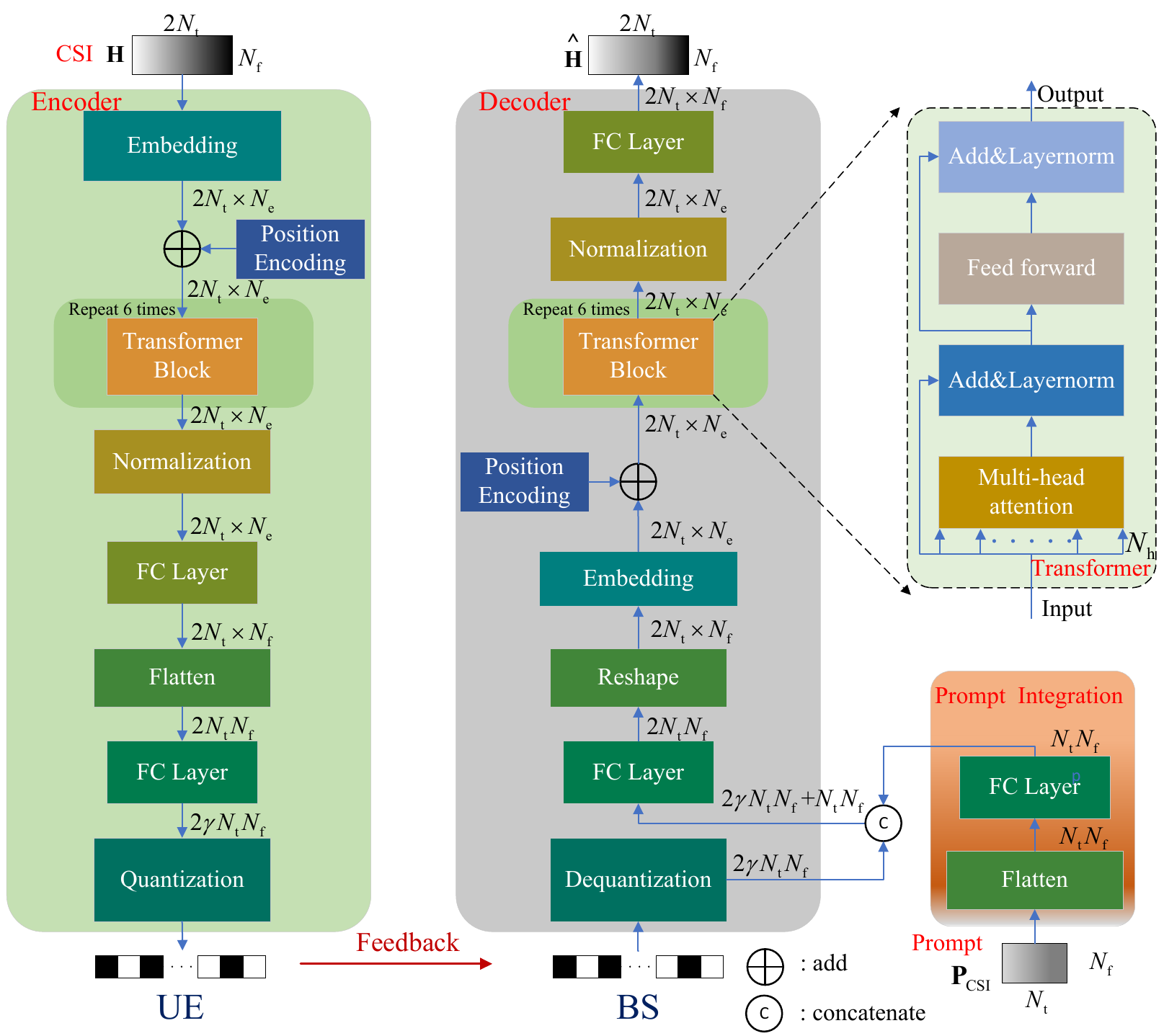}
    \caption{NN architecture of the transformer-based prompt-enabled LAM for CSI feedback. The encoder at the user compresses and quantizes downlink CSI ($\bf H$) to reduce feedback overhead, while the decoder at the BS reconstructs CSI ($\widehat{{\bf H}}$) using the feedback codeword and the prompt (${P}_{\rm CSI}$). Both encoder and decoder employ six transformer blocks to ensure robust performance.}
    \label{AEarchitecture} 
\end{figure*}

\subsection{Detailed NN Architecture}
 
Leveraging the exceptional performance of the transformer structure \cite{10229094,9961131}, this study adopts transformer blocks as the backbone for the CSI feedback LAM. The detailed architectures of the encoder, decoder, and transformer blocks are illustrated in Figure \ref{AEarchitecture} and described below. 

\subsubsection{Encoder Architecture}

As depicted in the left portion of Figure \ref{AEarchitecture}, the encoder module first extracts features from the original CSI matrix ($\mathbf{H}$), compresses these features into a low-dimensional vector, and then discretizes the compressed vector. Unlike (\ref{dftAD}), the input CSI matrix for the encoder ($\bf H$) has a dimension of ${2N_{\rm t} \times N_{\rm f}}$. This adjustment is necessary because NN libraries, such as TensorFlow, cannot directly process complex-valued matrices. Therefore, the real and imaginary components of the complex CSI matrix are stacked along the angular dimension.

Feature extraction is performed using six transformer blocks, preceded by an embedding layer with a dimension of $N_{\rm e}$ and sinusoidal position encoding. The extracted features are normalized, and their dimensions are restored via an FC layer to match the input dimension. A flatten layer is then used to reshape the extracted feature matrix into a one-dimensional vector. This vector undergoes compression through an FC layer with ${2\gamma N_{\rm t} N_{\rm f}}$ neurons, where $\gamma$ is the compression ratio. Finally, the compressed vector is quantized using a $b$-bit uniform quantizer.

\subsubsection{Decoder Architecture}

The decoder architecture, depicted in the central portion of Figure \ref{AEarchitecture}, incorporates a prompt-aided design. Unlike traditional decoders that reconstruct CSI solely based on the feedback codeword, this decoder reconstructs CSI using both the feedback codeword and the prompt (${\bf P}_{\rm CSI}$), which represents environmental knowledge.

In the prompt branch, the prompt (${\bf P}_{\rm CSI} \in \mathbb{R}^{N_{\rm f} \times N_{\rm t}}$) is first flattened into a one-dimensional vector and processed by an FC layer with $N_{\rm t} N_{\rm f}$ neurons. The preprocessed prompt is then concatenated with the feedback codeword, resulting in a combined vector of dimension $1\times (2\gamma N_{\rm t} N_{\rm f} + N_{\rm t} N_{\rm f})$. 

The combined vector is passed through an FC layer and a reshape layer to match the CSI input dimension. Similar to the encoder, the initially reconstructed CSI is processed through six transformer blocks, preceded by an embedding layer with a dimension of $N_{\rm e}$ and sinusoidal position encoding. The output of the final transformer block is normalized and adjusted in dimension through an FC layer, producing the final reconstructed CSI $\widehat{{\bf H}}$.

\subsubsection{Transformer Block}
The core architecture of the backbone transformer block is shown in the upper right part of Figure \ref{AEarchitecture}. The key component of the transformer block is the self-attention mechanism \cite{9961131}.
The input to this block is first processed by a multi-head attention layer with $N_{\rm h}$ heads, followed by a residual connection and layer normalization. Subsequently, the output undergoes a feedforward transformation implemented using two FC layers. After this transformation, another residual connection is applied, followed by layer normalization. The normalized output serves as the final output of the transformer block.
The complexity of the transformer block is primarily determined by factors such as the input dimension, the number of heads, the width of the FC layers, and related parameters. In this study, the width of the first FC layer is set to $4N_{\rm e}$, while the width of the second FC layer is specified as $N_{\rm e}$.

\section{Numerical Simulation and Discussions}
\label{s5}


\subsection{Simulation Settings}

\subsubsection{Channel Data Generation}
\label{CSIgenera}

To fully leverage the capabilities of the proposed prompt-enabled LAM across diverse wireless scenarios, we constructed a series of CSI datasets using the QuaDRiGa channel generator \cite{6758357}, adhering to 3GPP standards.
During channel generation, the carrier frequency is set to 2 GHz, employing the 3GPP\_38.901\_UMi\_NLOS channel model \cite{3gpp2018study}. The BS and the user are equipped with a ULA featuring $N_{\rm t}=32$ elements and a single receiving antenna, respectively. The bandwidth ($B$) is configured at 10 MHz, with $N_{\rm f} = 64$ subcarriers.
The number of clusters ($L$) is set to 4, each comprising 10 sub-paths.

In each scenario, users are positioned within a circular area with a radius ($R$) of 5 m, randomly placed within a cell. The scatterer distribution within each scenario's cell is entirely fresh, ensuring variability even at fixed locations across scenarios. A total of 3,200 scenarios are considered, with channels for 400 users generated in each scenario, resulting in 1,280,000 total channels. This significantly exceeds the sample sizes in previous studies, such as 105,996 in \cite{10677401}, 216,000 in \cite{liu2024wifowirelessfoundationmodel}, and 820,000 in \cite{LWM2024}.

After conversion into the angular-delay domain using (\ref{dftAD}), as described in \cite{wen2018deep}, we retain the initial 32 rows of ${\bf H}$, discarding the remaining 32 rows due to the restricted time-delay range of multipath arrivals. The resulting truncated CSI matrix has dimensions of $32\times 32$. For simplicity, we continue referring to this truncated matrix as ${\bf H}$.

\subsubsection{NN Training Details}

The training of CSI feedback LAMs is conducted on an Nvidia DGX-2 workstation using TensorFlow (Version: 2.14.0). All NNs, including LAMs with and without prompts, are trained end-to-end. The head number ($N_{\rm h}$), embedding dimension $N_{\rm e}$, and quantization bit number ($b$) are set to 16, 128, and 4, respectively. The NN training batch size, epoch count, and learning rate are configured at 256, 100, and 0.001, respectively. The Adam optimizer is used to update NN parameters.

Out of the 3,200 scenarios generated, 3,000 are randomly selected for training the feedback NNs. Specifically, 350 samples are chosen randomly within each scenario for training, while the remaining samples are reserved for evaluation. The prompt (${\bf P}_{\rm CSI}$) for each scenario is computed from these 350 CSI samples. During training, 10\% of the training samples are set aside for validation, and the NNs exhibiting the best performance on this validation set are retained as the final feedback NNs.

\begin{figure*}[t]
 \centering
\subfigure[3,000 seen scenarios.] {
 \label{largeSmallSeen}
\includegraphics[width=0.46\linewidth]{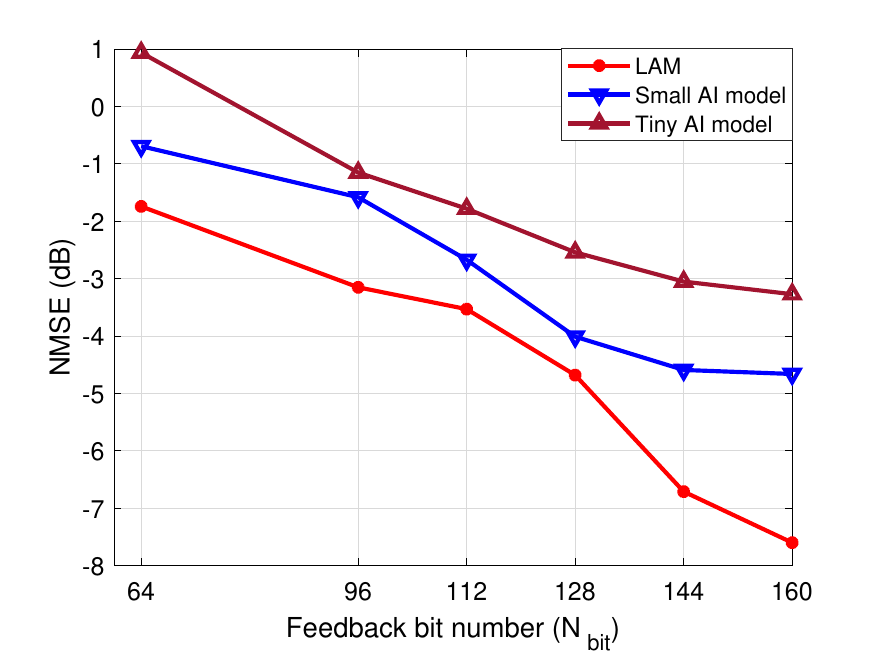}
}
\subfigure[200 unseen scenarios.] {
\label{largeSmallUnseen}
\includegraphics[width=0.46\linewidth]{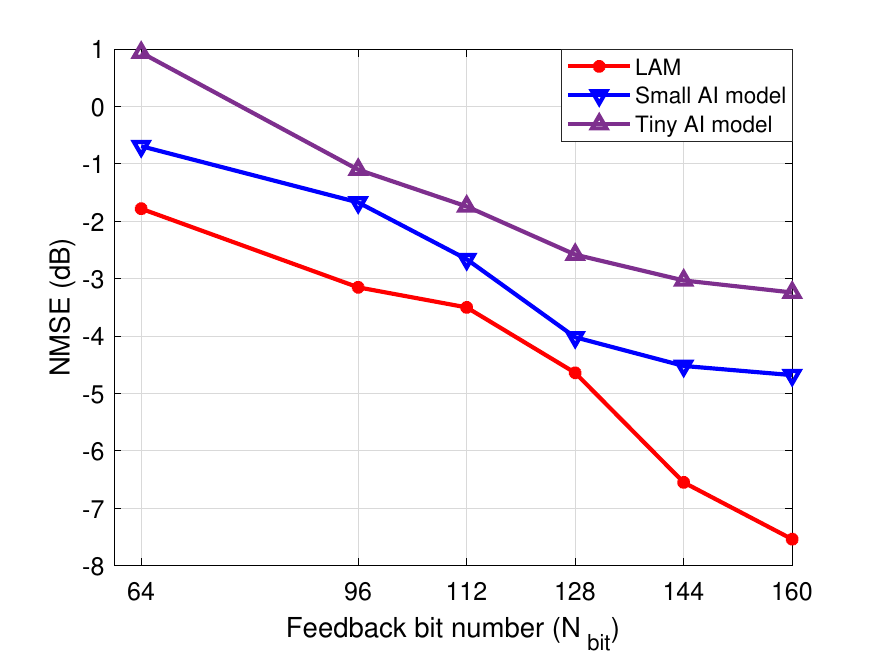}
}
\caption{NMSE performance comparison between the proposed CSI feedback LAM and small-scale AI models across varying feedback bit numbers ($N_{\rm bit}$) in 3,000 seen and 200 unseen scenarios. Regardless of the feedback bit number, the feedback LAM consistently outperforms the two baseline AI models. Moreover, all models demonstrate almost identical performance in both seen and unseen scenarios, highlighting their remarkable generalization abilities.}
\label{largeSmall}
\end{figure*}

\subsubsection{Baselines}

To evaluate the effectiveness of the proposed prompt-enabled feedback LAM, which integrates a larger, more powerful model and utilizes environmental knowledge, we compare it with several AI-based baseline frameworks: 

\begin{itemize}
\item {\bf Tiny AI Model for CSI Feedback:} This model's encoder consists solely of FC layers, excluding transformer blocks and preprocessing steps such as embedding and positional encoding, compared to the proposed LAM. In the decoder, only one transformer block is used to enhance the initially reconstructed CSI. Prompts are not used in this framework. Consequently, the NN complexity is significantly lower than that of the LAM shown in Figure \ref{AEarchitecture}.

\item {\bf Small AI Model for CSI Feedback:} This framework uses a single transformer block in both the encoder and decoder, significantly reducing NN complexity compared to the proposed LAM. Other components, such as preprocessing, remain the same as in the proposed LAM. Prompts are not utilized in this framework.

\item {\bf LAM Without Prompts for CSI Feedback:} This baseline shares the same NN architecture as the proposed model (six transformer blocks in both the encoder and decoder) but excludes the prompt branch shown in the lower right part of Figure \ref{AEarchitecture}.
 
\end{itemize}

\subsection{Results and Analysis}
In this subsection, we first compare the feedback LAM with smaller-scale feedback models, followed by an evaluation of the performance differences between LAMs with and without prompts. 

\begin{table}[t]
\centering
\caption{NN parameter numbers of different feedback AI models.}
\label{NNparameters}
\begin{tabular}{ccccc}
\toprule
\diagbox{\textbf{Model}}{$\mathbf{N}_{\rm bit} $} & \textbf{64} & \textbf{96} & \textbf{112} & \textbf{128} \\
\hline
\textbf{LAM} & 2,464,080 & 2,496,856 & 2,513,244 & 2,529,632 \\
\textbf{Small AI Model} & 481,360 & 514,136 & 530,524 & 546,912 \\
\textbf{Tiny AI Model} & 275,536 & 308,312 & 324,700 & 341,088 \\
\bottomrule
\end{tabular}
\end{table}

\subsubsection{Comparison Between Feedback LAM and Small-scale Models} 

Table \ref{NNparameters} presents the number of NN parameters for the CSI feedback LAM and small-scale AI models. As previously mentioned, the feedback LAM incorporates a significantly greater number of transformers in both the encoder and decoder compared to the small-scale feedback models. Consequently, the parameter count of the feedback LAM surpasses that of the smaller models. This indicates that the feedback LAM is expected to possess a more powerful fitting capability, which will be evaluated in the subsequent section.
 
Figure \ref{largeSmall} compares the NMSE performance across varying feedback bit numbers ($N_{\rm bit}$) on the 3,000 seen and 200 unseen scenarios. On the 3,000 seen scenarios used during NN training, the proposed LAM outperforms the two baseline AI models. When the feedback bit number is large, the performance gap becomes more pronounced. The small AI model achieves better feedback accuracy compared to the tiny model due to the utilization of more transformer blocks in the decoder.
Furthermore, beyond a feedback bit number of 144, the performance improvement of small-scale AI models for CSI feedback plateaus. However, the feedback performance of the proposed LAM continues to improve with the feedback bit number, demonstrating that the LAM can effectively leverage feedback bit information to support high-quality feedback reconstruction. Similar trends are observed in the performance of different models on unseen scenarios, as shown in Figure \ref{largeSmallUnseen}.

In existing literature, a prevalent consensus is that if feedback NNs are not trained on a specific scenario, there will be a notable decline in their performance. For instance, as illustrated in \cite[Figure 4]{10622316}, the feedback NMSE value increases from approximately $-24$ dB to around $3.2$ dB when the feedback NN is not trained with data from the target (test) scenario.
However, compared with the feedback performance on the 3,000 seen scenarios depicted in Figure \ref{largeSmallSeen}, there is minimal performance degradation on the 200 unseen scenarios illustrated in Figure \ref{largeSmallUnseen}. For example, with the feedback bit number ($N_{\rm bit}$) set to 144, the NMSE values for the seen and unseen scenarios are $-6.71$ dB and $-6.55$ dB, respectively, showcasing a mere 0.16 dB accuracy difference.

Moreover, we trained feedback LAMs exclusively using CSI data from 30 and 300 randomly selected scenarios. When the feedback bit number is set to 144, the feedback NMSE values of these two trained LAMs on the 200 unseen scenarios are $0.60$ dB and $-3.37$ dB, respectively (these results are not included in the figure), demonstrating a significant performance decrease. This observation highlights that the feedback LAM exhibits a high generalization capacity when trained on more diverse datasets, effectively addressing a primary challenge in AI-based CSI feedback---namely, generalization \cite{9970357,3gpp843,10288574}.

\begin{figure*}[t]
 \centering
\subfigure[3,000 seen scenarios.] {
 \label{promptSeen}
\includegraphics[width=0.46\linewidth]{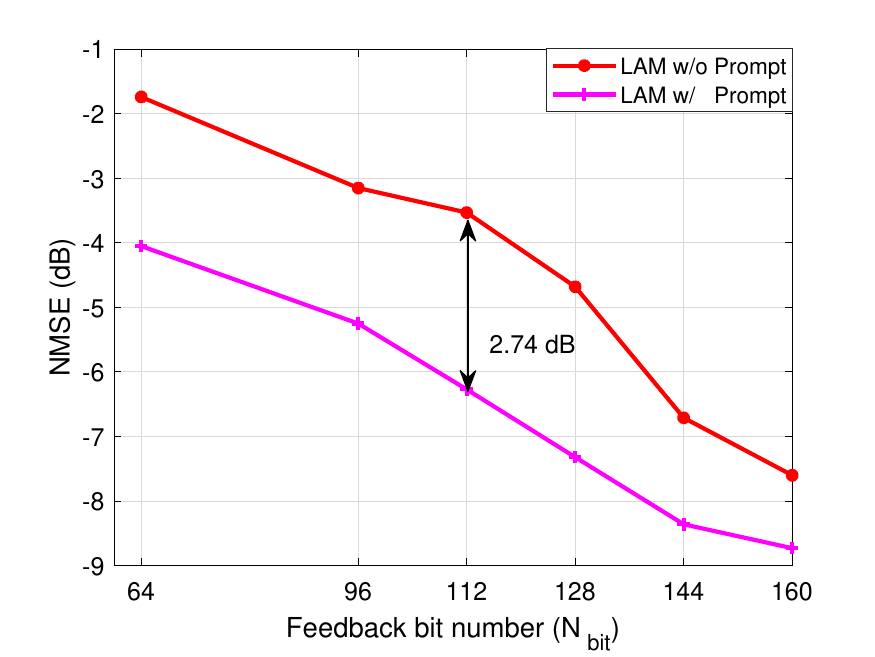}
}
\subfigure[200 unseen scenarios.] {
\label{promptUnseen}
\includegraphics[width=0.46\linewidth]{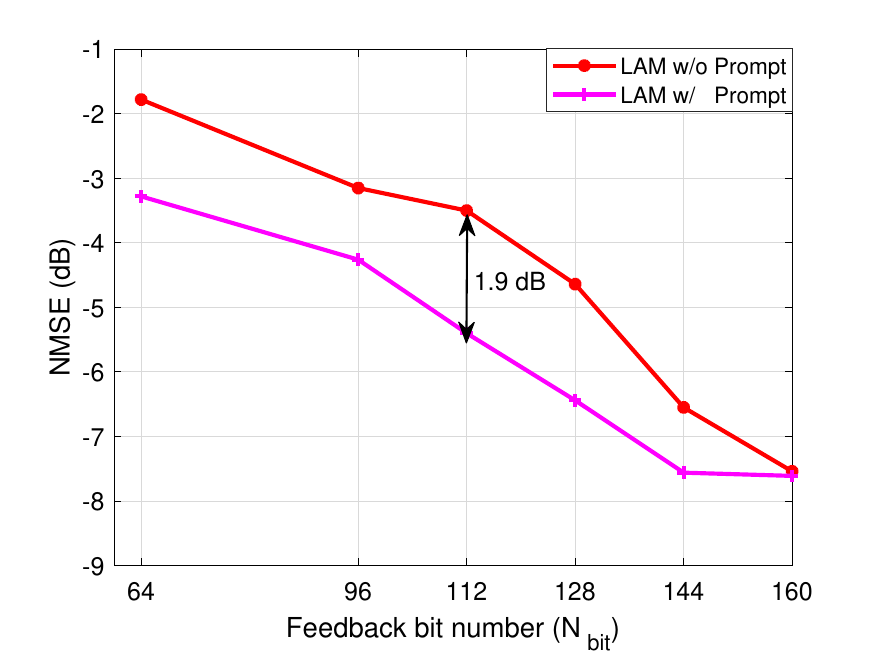}
}
\caption{NMSE performance comparison between the LAMs with and without the prompts across varying feedback bit numbers ($N_{\rm bit}$) on the 3,000 seen and 200 unseen scenarios. The utilization of the prompt, that is, environmental knowledge, considerably improves CSI feedback accuracy.}
\label{promptResult}
\end{figure*}
 
\subsubsection{Comparison between Feedback LAMs with and without Prompt}

Figure \ref{promptResult} compares the NMSE performance of the CSI feedback LAMs with and without prompts across varying feedback bit numbers ($N_{\rm bit}$) on 3,000 seen and 200 unseen scenarios. Firstly, regardless of whether the scenarios are included in LAM training, the incorporation of the prompt (${\bf P}_{\rm CSI}$) significantly enhances CSI feedback performance, demonstrating the superior capabilities of the proposed prompt-enabled LAM. Secondly, the degree of performance enhancement diminishes as the feedback bit number ($N_{\rm bit}$) increases. For example, in Figure \ref{promptUnseen}, with a feedback bit number of 112, the NMSE value improves by 1.9 dB. However, with a bit number of 160, the LAMs with and without prompts exhibit nearly identical performance.
The increase in feedback bits allows the decoder to acquire more information useful for downlink CSI reconstruction from the feedback codeword ($\bf s$), thereby reducing its reliance on the prompt.

On the other hand, the performance improvement from incorporating the prompt in unseen scenarios is significantly less compared to scenarios without prompt integration. For example, with 112 feedback bits, the NMSE improvements are 1.9 dB and 2.74 dB, respectively. This observation stems from the limited training data available for the prompt. Despite the training dataset comprising over 1,000,000 CSI samples, only 3,000 prompt samples are utilized for training the prompt-enabled LAM, resulting in a substantial performance disparity between seen and unseen scenarios. However, by including more scenarios during LAM training, this performance gap can be reduced.

\begin{figure}[t]
    \centering
    \includegraphics[width=0.92\linewidth]{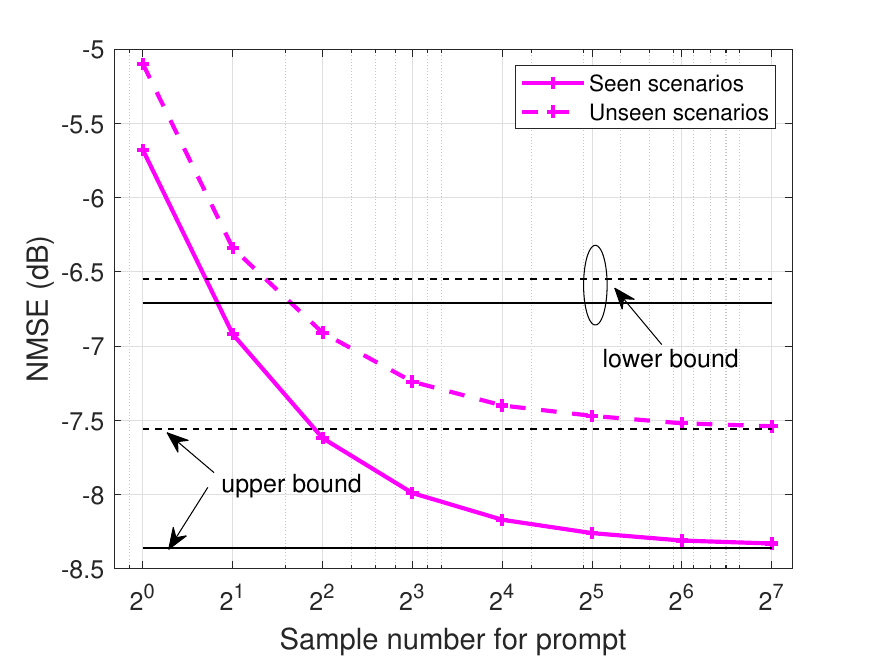}
    \caption{NMSE performance of the prompt-enabled LAM versus the number of samples for prompt generation on the 3,000 seen and 200 unseen scenarios. When the number of samples for prompt generation exceeds 32, the feedback LAM achieves performance comparable to the upper bound.} 
    \label{promptNumber}
\end{figure}

Although the LAM no longer requires online training when a new scenario emerges, prompt (${\bf P}_{\rm CSI}$) generation, as illustrated in (\ref{CSImagnitude}), necessitates the collection of CSI samples. Figure \ref{promptNumber} illustrates the NMSE performance of the prompt-enabled LAM across varying numbers of samples for prompt generation in both seen and unseen scenarios.
The lower bound represents the feedback performance of the LAM without a prompt, whereas the upper bound signifies the performance of the prompt-enhanced LAM, with the prompt generated from all 350 training CSI samples. When the number of samples for prompt generation exceeds 32, the feedback LAM achieves performance comparable to the upper bound.
This result stems from the fact that the prompt primarily describes the channel distribution of a given scenario, providing coarser information compared to individual CSI samples, and can thus be derived from a limited number of CSI samples. Hence, users do not need to store a large quantity of CSI samples to generate the prompt.
Additionally, in this part, the CSI samples for prompt generation are randomly chosen. If the CSI samples are not randomly selected, such as being collected at a fixed or nearby position, increasing the number of CSI samples will not improve performance since the channel characteristics of a specific position are stable.

\section{Conclusion and Future Directions}
\label{s6}
In this study, we introduced a prompt-enabled LAM for CSI feedback. To the best of our knowledge, this is the first work to apply LAMs to CSI feedback. Instead of aggressively stacking NN layers, our research distinguishes itself by probing the underlying mechanism of AI-based feedback, laying the foundation for integrating LAMs into the feedback process and introducing a prompt mechanism to enhance feedback LAMs.
Specifically, we initiated our investigation by exploring the feedback mechanism through various CSI datasets. Our findings suggested that AI-based CSI feedback heavily relies on the powerful fitting capabilities of NNs and the effective utilization of environmental knowledge. Building upon these insights, we focused on incorporating powerful LAMs and environmental knowledge effectively. The environmental knowledge was represented as prompts, integrated into the feedback LAM. Subsequently, we proposed a prompt-enabled LAM framework for CSI feedback, where both the encoder and decoder are based on powerful transformer blocks. Simulation results validate that leveraging LAMs not only significantly reduces feedback errors compared to small-scale AI models but also enhances feedback generalization capabilities. Moreover, the integration of prompts further improves feedback accuracy while minimizing the overhead associated with dataset collection for new scenarios.

Although the proposed prompt-enabled feedback LAM demonstrates promising results, several significant challenges warrant further in-depth exploration in the future:
\begin{itemize}
    \item {\bf LAM Architecture Design:} The feedback LAM architecture is directly adopted from computer science, disregarding the physical significance of CSI. Utilizing an advanced NN architecture tailored for CSI can enhance the performance of feedback LAMs.
    \item {\bf Complexity Reduction:} As shown in Table \ref{NNparameters}, the complexity of the LAM is considerably higher than that of small-scale AI models, posing a significant challenge to the deployment of AI algorithms in the air interface. Advanced NN compression techniques, such as knowledge distillation, present a potential solution \cite{tacl_a_00704}.

    \item {\bf Enhanced Prompt Design:} The prompt design in this study is relatively basic, and improving it could lead to better performance \cite{van2023chatgpt}.

\end{itemize}

\bibliographystyle{IEEEtran}
\bibliography{reference}

\end{document}